\documentclass[epj-spec]{svjour}
\usepackage{graphicx}
\newcommand{\et}{\textit{et al. }}
\newcommand{\hhh}{\mbox{H$_2$ }}
\newcommand{\grad}{\mbox{\(\mathsurround=0pt{}^\circ\)}}
\newcommand{\kms}{\mbox{$\mathrm{km\,s^{-1}}$}}
\newcommand{\HI}{\mbox{H\textsc{i}}\,\,}
\begin{document}
\title{Variability of the proton-to-electron mass ratio on cosmological
scales}
\author{M.Wendt\inst{1}\fnmsep\thanks{\email{mwendt@hs.uni-hamburg.de}} \and
D.Reimers\inst{1} }
\institute{Hamburger Sternwarte, Universitat Hamburg, Gojenbergsweg 112,
21209 Hamburg, Germany}
\abstract{
The search for a possible variation of fundamental physical constants is
newsworthy more than ever. A multitude of methods were developed. So far the
only seemingly significant indication of a cosmological variation exists
for the proton-to-electron mass
ratio as stated by Reinhold \et \cite{Reinhold06}. The measured
indication of variation  is based on  the combined analysis of \hhh absorption
systems in the spectra of
\texttt{Q0405-443} and \texttt{Q0347-383} at $z_{\mathrm{abs}}=2.595$ and 
$z_{\mathrm{abs}}=3.025$, respectively. The high resolution data of the latter
is reanalyzed in this work to examine the influence of different fitting
procedures and further potential nonconformities. This analysis cannot
reproduce the significance achieved by the previous works.
} 
\maketitle
\section{Introduction}
\label{intro}
Contemporary theories of fundamental interactions, particularly those subsumed
as string theories allow for all kinds of variations of fundamental constants in
the course
of the evolution of the universe. Variations of the coupling constants of
strong and electroweak interactions would affect the masses of elementary
particles dependent on the model for the expanding universe.\\
It is up to cosmology to achieve first
estimates and constraints of possible variations on large scale since laboratory
experiments lack the spatial or temporal coverage.
A recent analysis of HE 0515-4414 by Chand \et \cite{Chand06} and Molaro \et
 \cite{Molaro07} yielded a result for the fine structure constant that is in
good agreement with
no variation at the current level of accuracy. However there are
results indicating a variation in $\alpha$ or not yet fully controlled
systematics (Murphy \et \cite{Murphy04}, Levshakov \et \cite{Levshakov07}).
A measured trend of positive variation in the proton-to-electron mass
ratio by Ivanchik \et \cite{Ivanchik05} and a later refinement to
$\Delta\mu/\mu=(2.0\pm0.6)\times10^{-5}$ \cite{Reinhold06} launched numerous
theories with
different scenarios to interpret the new data. The subject is still under heavy
debate and confirmation of the results and methods is required.
\section{Detecting $\Delta\mu/\mu$ via molecular hydrogen}
\label{sec:1}
\begin{figure}[h]
\begin{tabular}{cc}
\includegraphics[width={0.5 \columnwidth}]{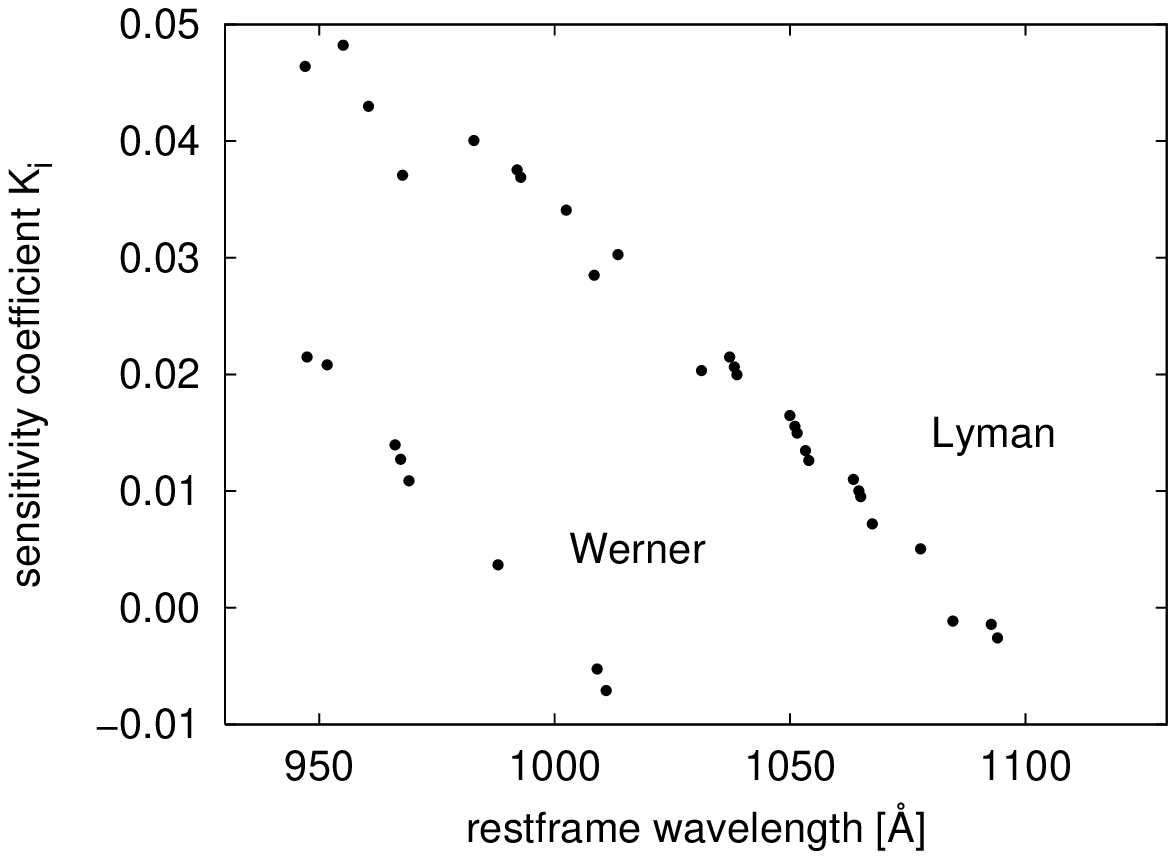}&
\includegraphics[width={0.5 \columnwidth}]{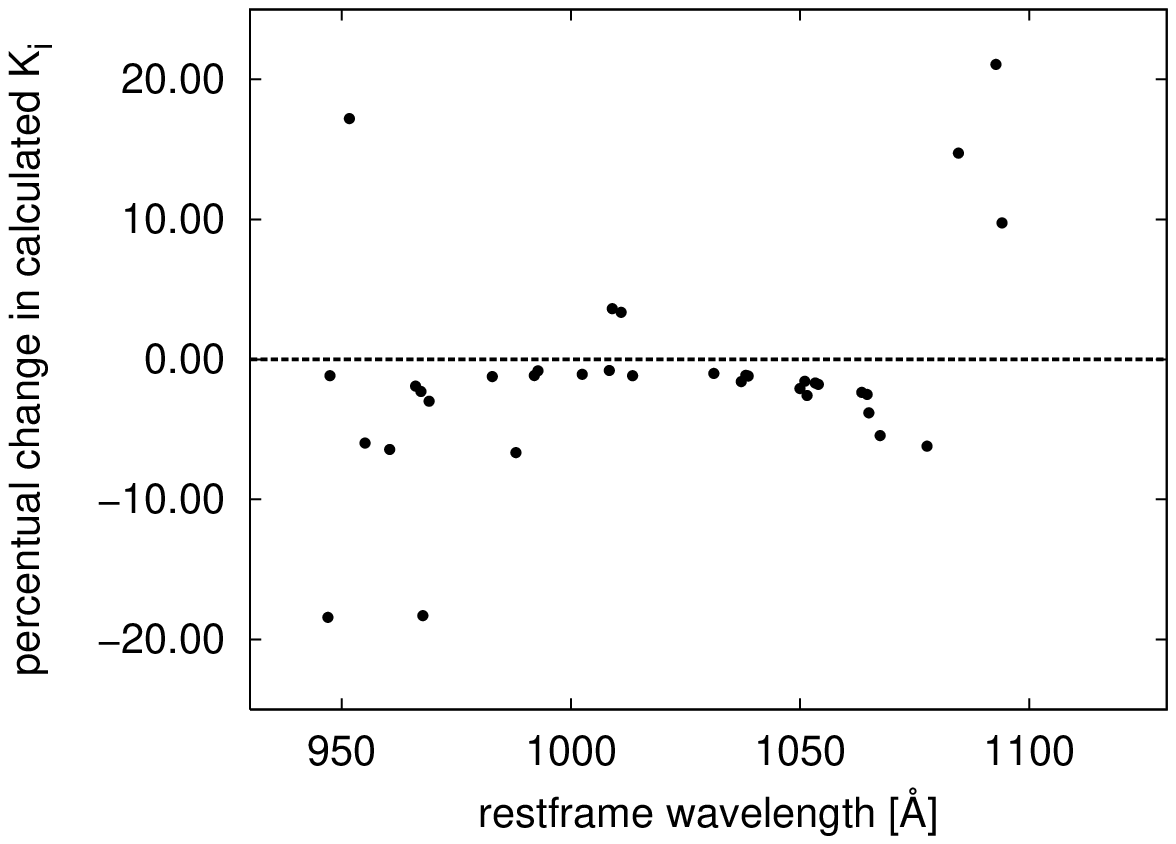}\\
\end{tabular}
\caption{Sensitivity coefficients of observed lines in the Lyman and Werner band
(\textit{left}) and recent refinement of $K_i$ given in percentages
(\textit{right}).}
\label{fig:ki_plot}
\end{figure}

The latest measured value of the proton-to-electron mass ratio is
$\mu_0 = 1836.15267261(85)$ 
\cite{Mohr05}. The precision reached by today's laboratory
experiments rules out considerable variation of $\mu$ on short time scales but
cannot yet exclude a change over cosmological timescales on the order of
$10^{10}$ years. Additionally the possibility of different proton-to-electron
mass ratios in widely separated regions of the universe cannot be
rejected. 
The method to constrain a possible variation of $\mu$ as applied in this work
was first suggested by Thompson \cite{Thompson75}.
Electronic, vibrational, and rotational excitations of a diatomic molecule
depend differently on its reduced
mass $\mu$. 
To a first approximation, these energies are
proportional to $\mu^0$, $\mu^{-\frac{1}{2}}$, $\mu^{-1}$, respectively.
Hence each transition has an individual sensitivity to a possible change in
that reduced mass. This can be expressed by a sensitivity coefficient.
\begin{equation}\label{eq:ki}
K_i=\frac{\mathrm{d}\,\mathrm{ln}\,\lambda_i}{\mathrm{d}\,
\mathrm{ln}\,\mu}=\frac{\mu}{\lambda_i}\frac{
\mathrm{d}\lambda_i}{\mathrm{d}\mu}.
\end{equation}
The calculations of $K_i$ were recently refined by Reinhold
\et \cite{Reinhold06}.
In first order they can be expressed by the Dunham coefficients $Y_{kl}$ of the
ground and excited states. With $\mu_n=\frac{m_e\mu}{2}$, Equation \ref{eq:ki}
leads to:
\begin{equation}\label{eq:ki_eq}
K_i
=-\frac{\mu_n}{\lambda_i}\frac{\mathrm{d}\lambda_i}{\mathrm{d}\mu_\mathrm{n}}
=\frac{1}{
E_\mathrm{e}-E_\mathrm{g}}
\left(-\frac{\mu_\mathrm{n}\mathrm{d}E_\mathrm{e}}{\mathrm{d}\mu_\mathrm{n}}
+\frac{ \mu_\mathrm{n}\mathrm{d} E_\mathrm{g}}{
\mathrm{ d}\mu_\mathrm{n}} \right).
\end{equation}
The computations for the energies of the excited and ground state, $E_e$ and
$E_g$, respectively are the same as for the energy levels of H$_2$.
Starting with the BOA based on the
semiempirical approach the energy levels can be expressed by the Dunham formula
\begin{equation}
E(v,J)=\sum_{k,l}Y_{kl}\left(v+\frac{1}{2}\right)^k[
J(J+1)-\Lambda^2]^l; \hspace{1cm}\Lambda^2=0 \mathrm{\hspace{0.3cm} (Lyman)}, 
\,1\mathrm{\hspace{0.3cm} (Werner)}
\end{equation}
However, the Dunham coefficients $Y_{kl}$ cannot be calculated directly from
the level energies due to strong mutual interaction between the excited states
as well as avoided rotational transitions between nearby vibrational levels.
For the first time the more complex non-BOA effects are taken into account
in the work of Reinhold \textit{et al. }\cite{Reinhold06}. Their recent
precise laboratory measurements of the level energies of
molecular hydrogen allowed for a reliable enhancement of the BOA approximation.
The influence of the inclusion of the Bohr shift and adiabatic
corrections is shown in Figure \ref{fig:ki_plot} (\textit{right}) in percentage
compared to prior sensitivity coefficients \cite{Varshalovich95}.
\subsection{Distinguish cosmological redshifts from variation of $\mu$}
With $K_i$ from Equation \ref{eq:ki}, the rest frame wavelengths
$\lambda_i^0$ are related to those in the quasar absorption system $\lambda_i$
via
\begin{equation}
\lambda_i=\lambda_i^0(1+z_{\mathrm{abs}})\left(1+K_i\frac{\Delta
\mu}{\mu}\right),
\end{equation}
with $z_{\mathrm{abs}}$ as the redshift of the absorbing system.
This can be expressed in terms of the individual redshift of each measured
\hhh component:
\begin{equation}\label{eq:linear_plot}
z_i=z_{\mathrm{abs}}+b K_i.
\end{equation}
According to Equation \ref{eq:linear_plot} the redshift of each \hhh feature
can be distinguished between the intrinsic redshift of the absorber and an
additional
component due to a possible variation in $\mu$.
Equation \ref{eq:linear_plot} describes a simple linear equation with a gradient
of
\begin{equation}
b=(1+z_{\mathrm{abs}}) \frac{\Delta \mu}{\mu}.
\label{eq:zvsk_gradient}
\end{equation}
A linear regression of the redshift of each \hhh line $z_i$ and its individual
sensitivity coefficient $K_i$ will yield the redshift of the absorbing
DLA system and $\Delta \mu/\mu$ during the epoch between $z=z_{\mathrm{abs}}$
and today.
It is worth noting that in the case of non-zero variation the redshift of the
absorber is not identical to the mean redshift of all observed lines as can be
derived from Equation \ref{eq:linear_plot}.
%
\section{Observations}
The source (\texttt{Q0347-383}) of the analyzed spectrum is a bright
quasi-stellar radio object
(QSO) with a visual magnitude of $V=17.3\,\,\mathrm{mag}$ at a redshift of
$z=3.23$ \cite{Maoz93}. 
The precise position dated to 2000 as stated in the fifth fundamental catalogue
is $\alpha$ 03h 49m 43.68s,  $\delta$ -38\grad 10' 31.3''.\\
The Quasar absorption line spectra were obtained with the Ultraviolet and
Visual Echelle Spectrograph (UVES) at the Very
Large Telescope (VLT) of the European Southern Observatory (ESO) in Paranal,
Chile.
The slit was 0.8 arcsec wide resulting in a spectral resolution of $R$ $\approx$
53.000 over the wavelength range 3300\AA\, -- 4500\AA.\\
The average seeing during observation was about 1.2 arcsec. Before and after the
exposures for each night, Thorium-Argon calibration data were taken.
An overall of nine spectra were recorded with an exposure time of 4500 seconds
each between January 8th and January 10th 2002 for the ESO program
\texttt{68.A-0106(A)}. All spectra were taken with grating 430 and the blue
``Pavarotti''-CCD with 2x2 binning. Later on the data were reduced manually by
Mirka Dessauges-Zavadsky from Geneva Observatory in Jan 2004 to achieve
maximum accuracy. The ESO Ambient Conditions
Database\footnote{http://archive.eso.org/eso/ambient-database.html} includes
measurements of the environmental parameters at the Paranal ESO observatory and
shows no significant changes in temperatures during or inbetween the
exposures that could lead to shifts between the separate observations.
The reduced data used here is identical to the one in
\cite{Ivanchik05} and \cite{Reinhold06}.
\section{Data analysis}
The nine separate spectra are coadded only to identify H$_2$ lines with a
higher signal-to-noise ratio. The fitting is carried out simultaneously on
the separate spectra to avoid influences of slight shifts between the
datasets.
The observed spectral range of the data  covers
over 80 \hhh lines of which after careful selection and avoidance of blends
only 39 are taken into account for the analysis. The \hhh lines are mostly
noticeable by their narrow line profiles and generally low equivalent width.
The observed spectrum has an instrumental resolution of 5.6 \kms. That
corresponds to a temperature of more than 4000K for Doppler
broadened H$_2$ lines. A lower excitation temperature is to be expected for
intergalactic molecular Hydrogen. The observed width of the H$_2$ features can
thus be expected to be given by the spectral resolution.
\subsection{Fitting of \hhh features}
\begin{figure}
\begin{center}
\begin{tabular}{ccc}
\includegraphics[height=4cm,width=4cm]{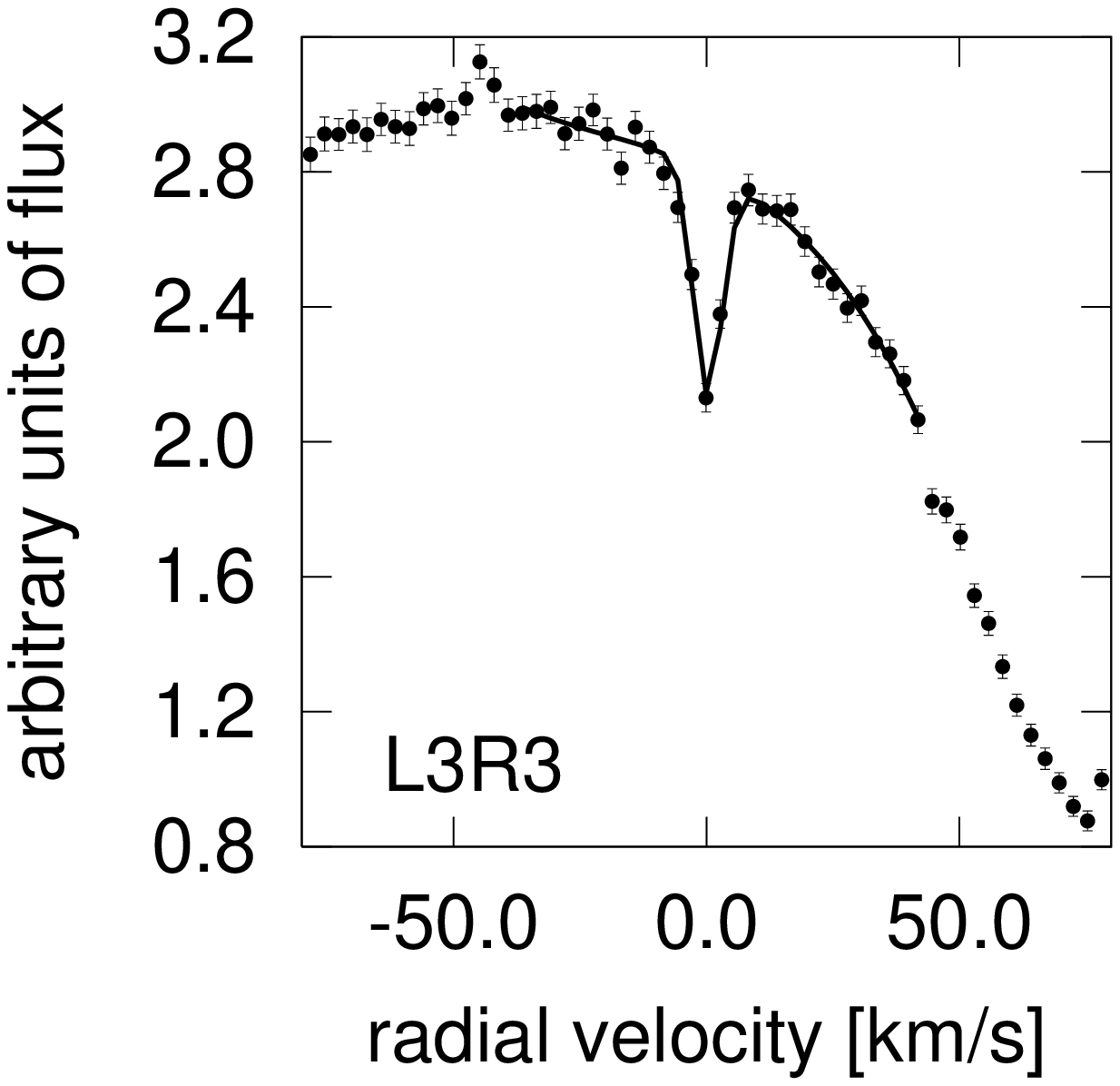} &
\includegraphics[height=4cm,width=4cm]{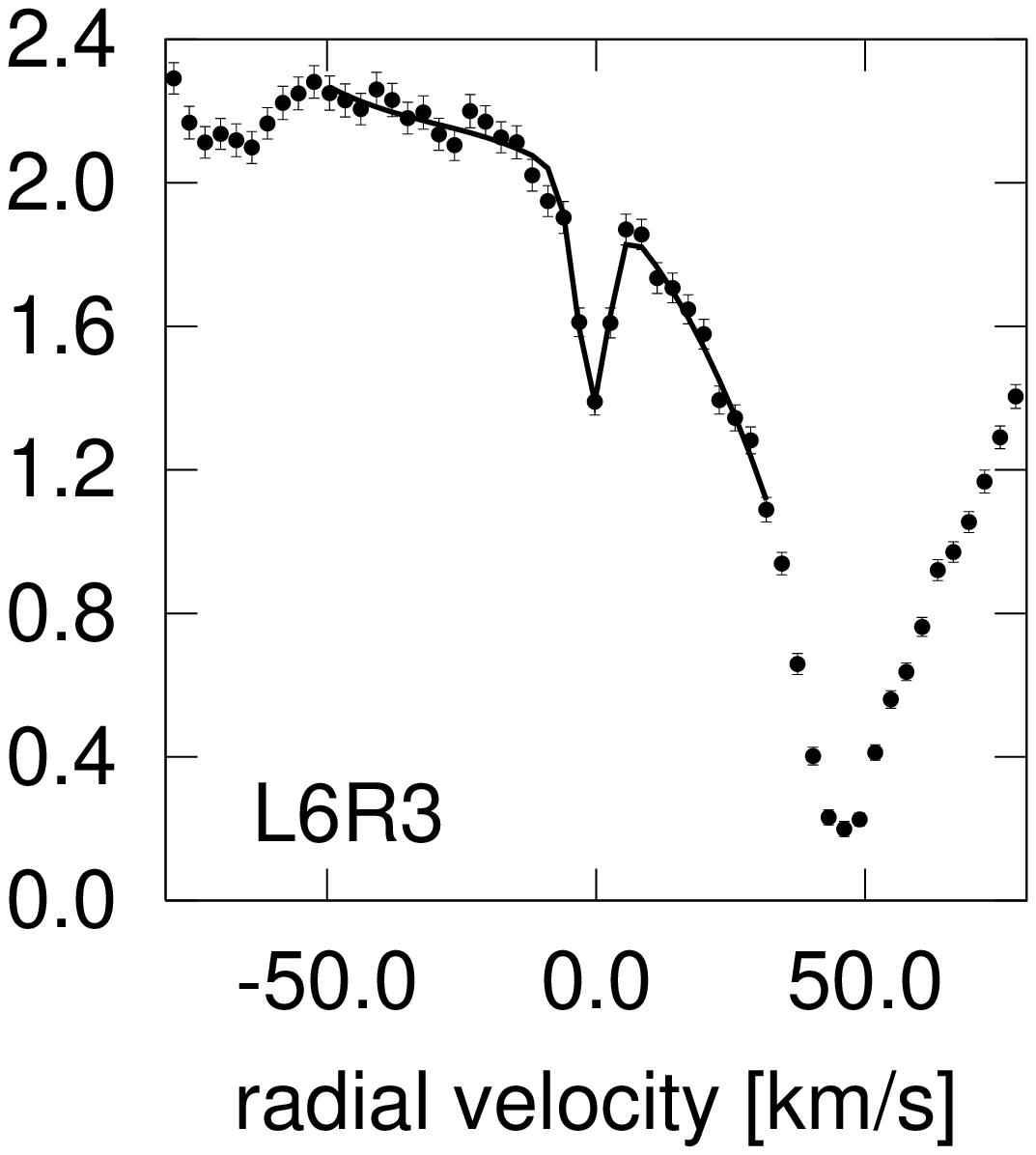} \\
\end{tabular}
\caption{Continuum matching via parabolic fit to the observed flux for
\texttt{L3R3} and \texttt{L6R3}.}
\label{fig:parabolic continuum}
\end{center}
\end{figure}

The selected and verified \hhh lines are fitted with an evolutionary fitting
algorithm \cite{Quast05}. Due to the nature of DLA systems the continuum is
heavily contaminated. 
The background is determined over a given range of the line feature and fitted
with a polynomial. That range is selected individually for each line by eye to
have  the best compromise between a clean continuum and the number of pixels
contributing to it. The grade of the polynomial is manually selected to match
the continuum flux. In some cases an additional line is fitted along with the
\hhh component to have a better fit to the flux. For those cases the true
continuum is fitted via a linear function. 
In cases of a continuum flux with apparent influences of broad Lyman $\alpha$
features or general contamination, a parabolic or cubic function is used to fit
the background to the observed flux. See Figure \ref{fig:parabolic continuum} 
of \texttt{L3R3} and \texttt{L6R3} for an exemplar of a parabolic fit.

\begin{figure}[b]
\begin{tabular}{cc}
\includegraphics[width={0.5 \columnwidth}]{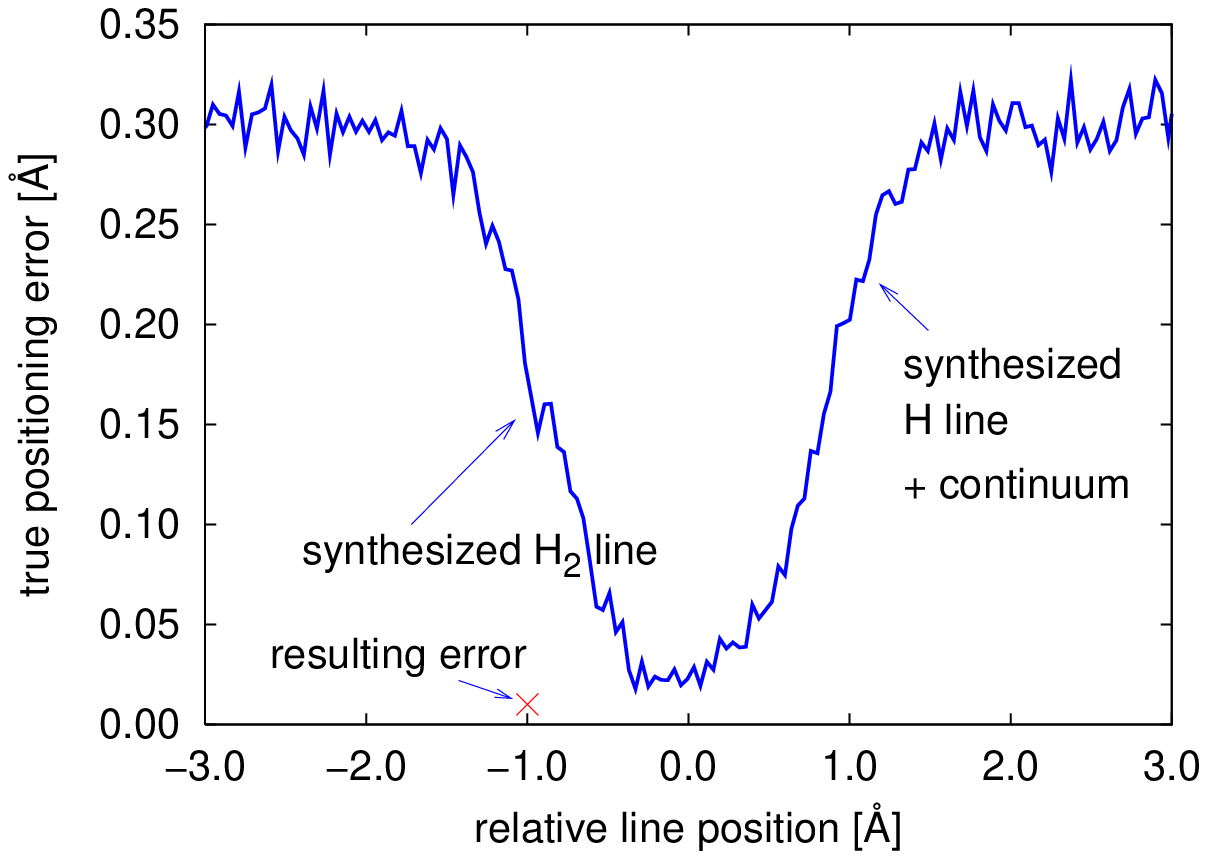}&
\includegraphics[width={0.5 \columnwidth}]{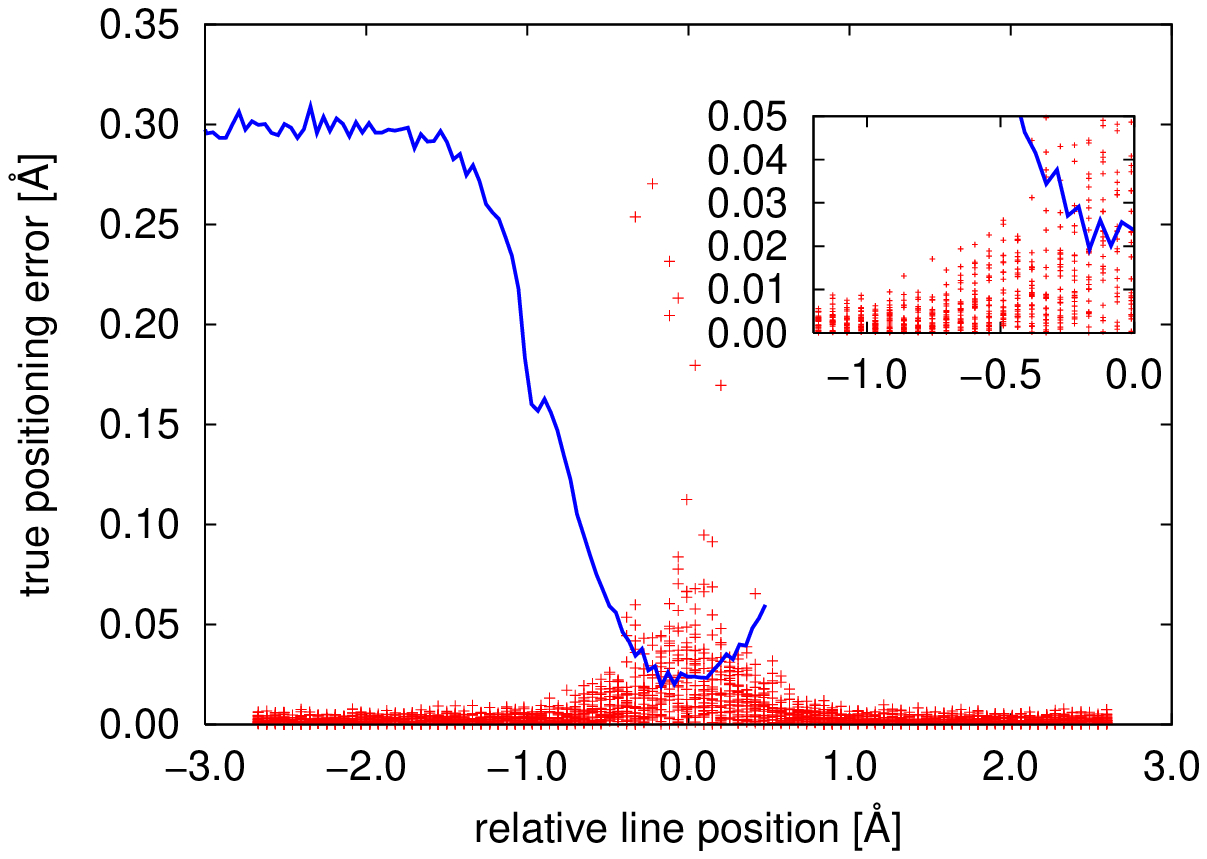}\\
\end{tabular}%
\caption{Series of fits of synthesized spectra. On the left an exemplified
case and on the right an overview with a blow up of the central region.}
\label{fig:concept}
\end{figure}

\begin{figure}
\begin{tabular}{cc}
\includegraphics[width={0.5 \columnwidth}]{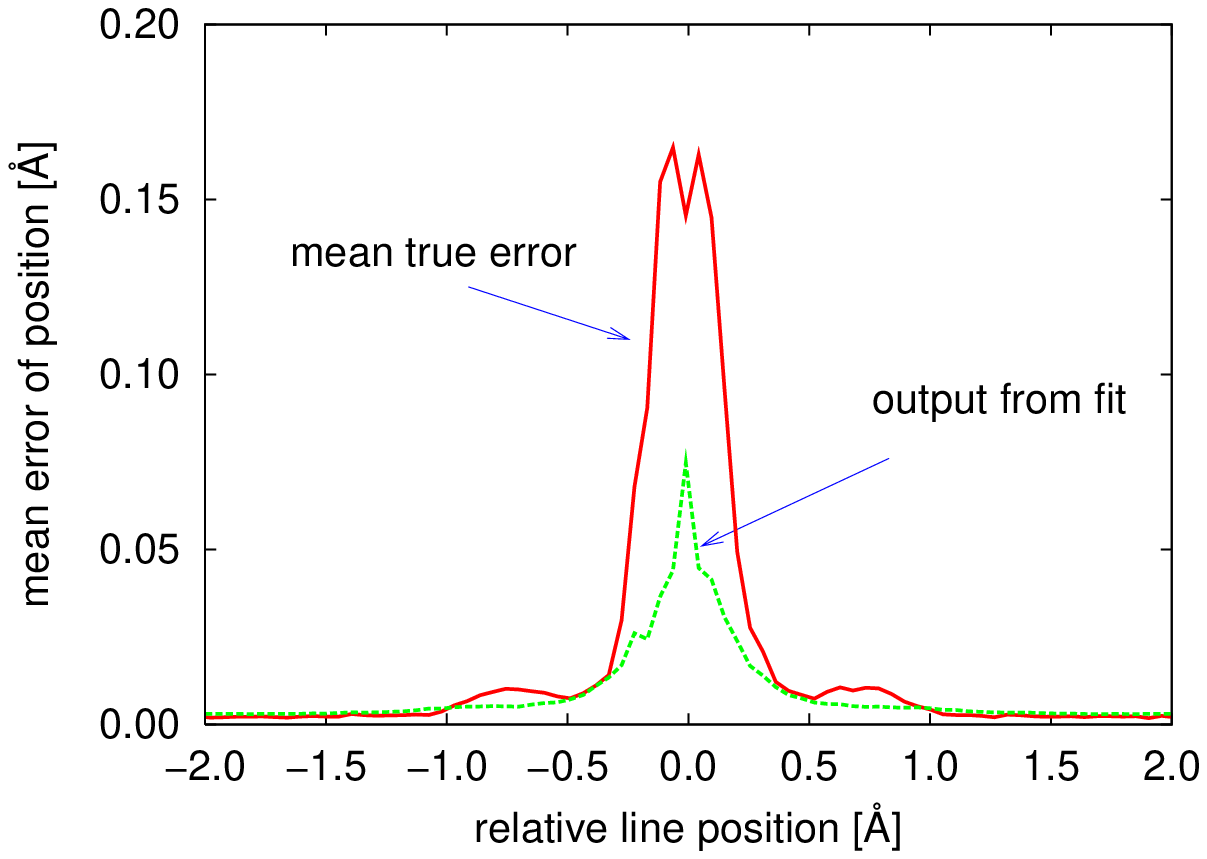}&
\includegraphics[width={0.5 \columnwidth}]{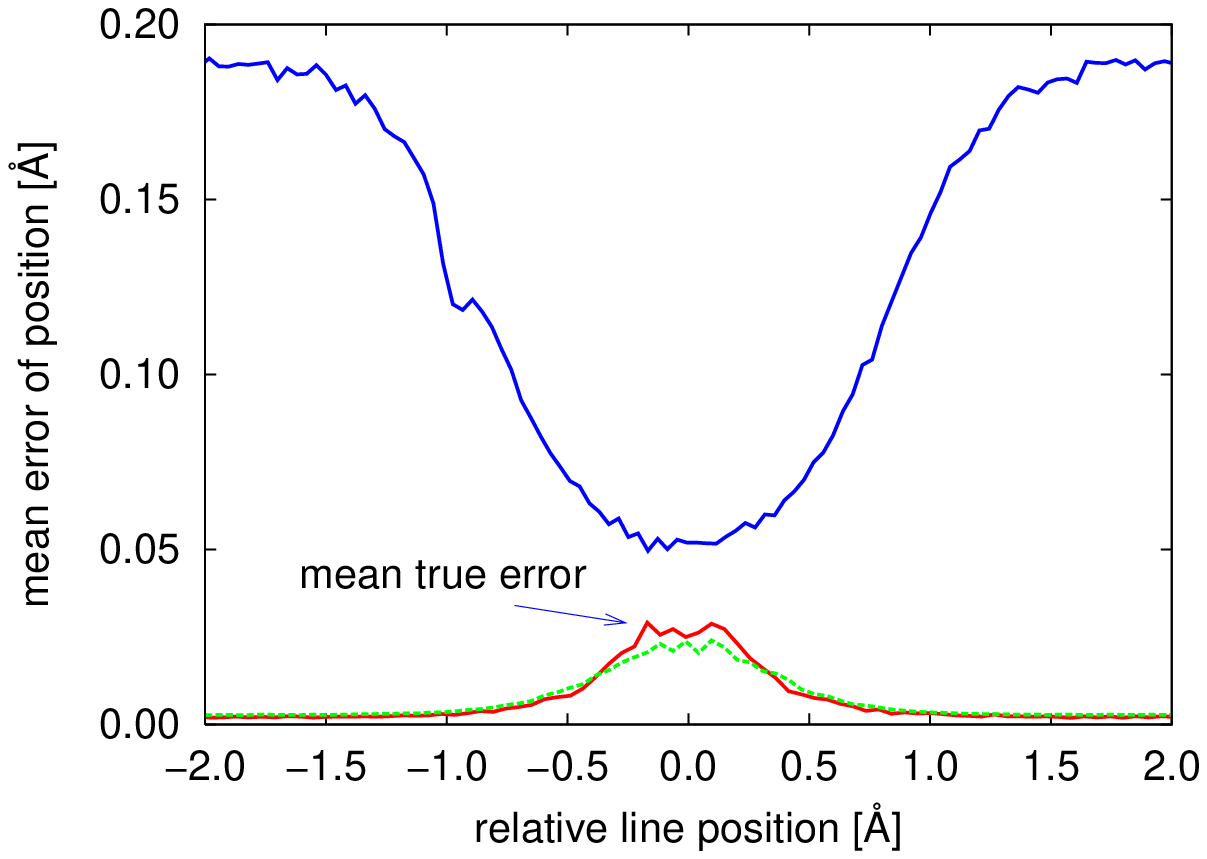}\\
\end{tabular}
\caption{Mean value of the true error and outcome of the fitting
procedure. Comparison of the single component fit (\textit{left}) and the two
component fit (\textit{right}).}
\label{fig:compare_sim_mean_err}
\end{figure}

\begin{figure}[b]
\begin{tabular}{cc}
\includegraphics[width={0.5 \columnwidth}]{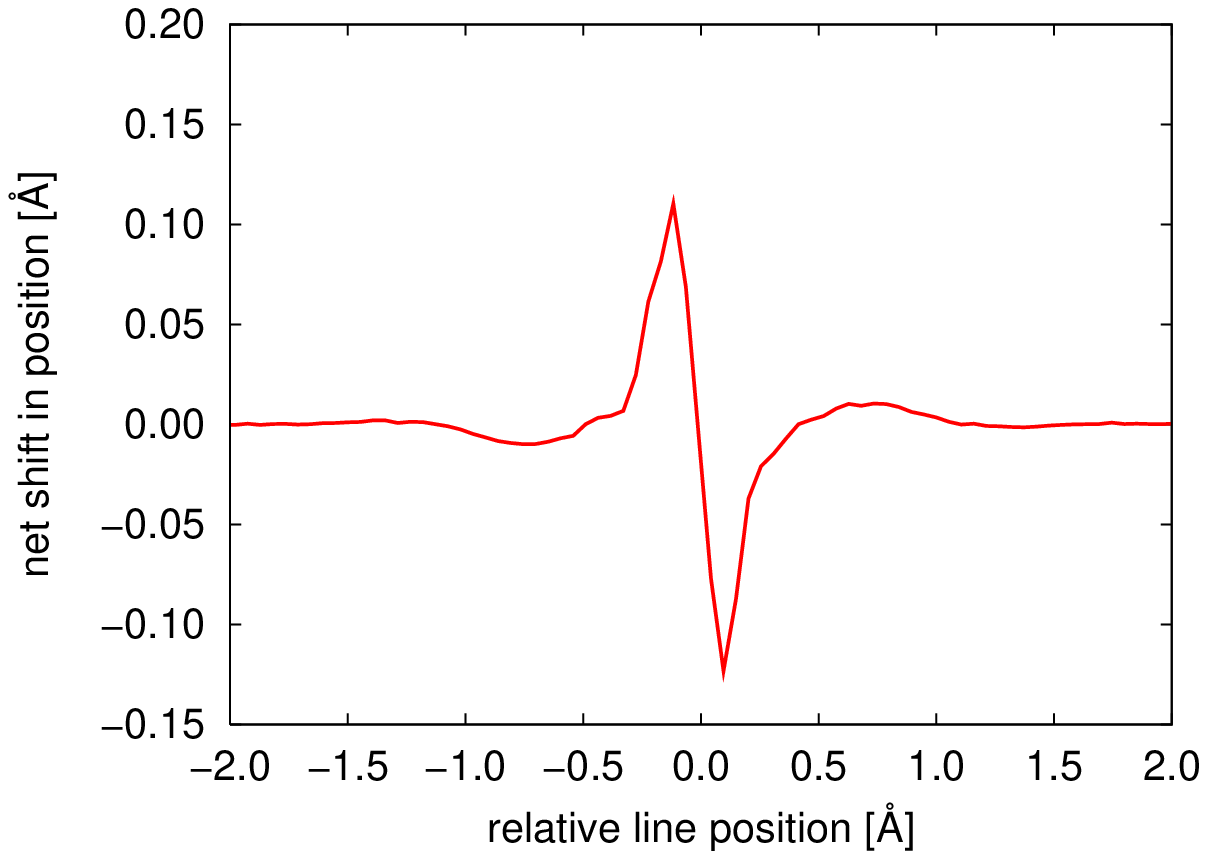}&
\includegraphics[width={0.5 \columnwidth}]{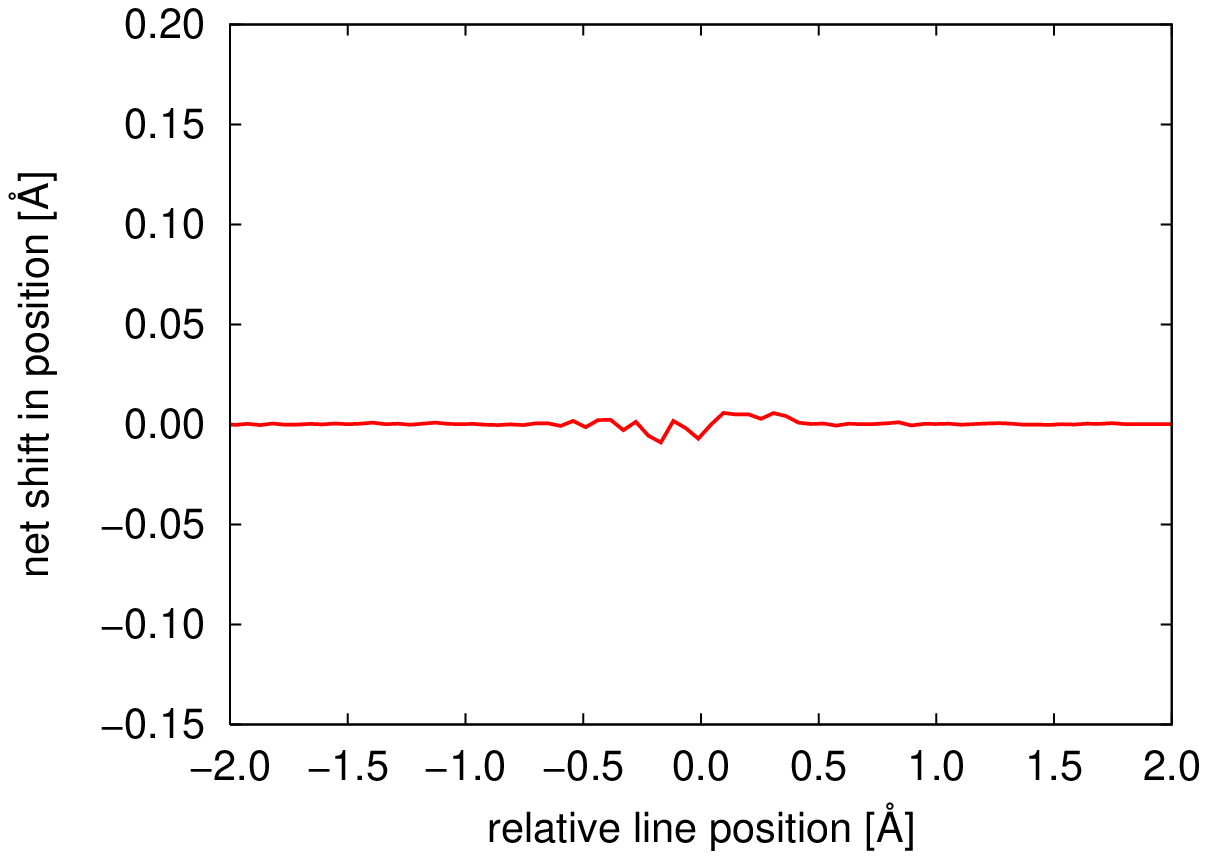}\\
\end{tabular}
\caption{Mean shifts of fitted position over a series of fits for the single
component fit (\textit{left}) and the two component fit (\textit{right}).}
\label{fig:compare_sim+shift}
\end{figure}

This approach -- however necessary -- is highly critical. The applied
rectification of the flux may have an impact on the fitted central position of
the line feature. For the analysis in \cite{Reinhold06} and \cite{Ivanchik05}
the estimated background is constructed manually.
To ascertain the validity of this approach a series of fits for simulated
spectra is carried out. A spectrum of a hypothetical \hhh feature on the
wing of a broad almost saturated \HI absorption feature is synthesized. Its
relative position to the center of the \HI feature is altered to cover the
full area of influence of the dominant atomic hydrogen feature. For each
constellation 100 spectra are synthesized and fitted. Figure
\ref{fig:concept} (\textit{left}) shows an example setup. For the given
position of the \hhh feature the resulting error in determining the center
of the \hhh absorption feature can be computed by the difference of the input
to the artificially generated spectrum and the output of the fitting algorithm.
On the \textit{right} panel of Figure \ref{fig:concept} the results of a full
run of 10,000 fits are plotted. 
Figure \ref{fig:compare_sim_mean_err} illustrates the averaged computed
absolute error and the error as ascertained by the fitting algorithm.
To evaluate the accuracy of a polynomial continuum fit, two separate runs are
carried out. One with a parabolic continuum fit (\textit{left} panel) and one
with a linear continuum but a second absorption line (in this case the broad \HI
line) as free parameter to match the observed flux (\textit{right} panel).
The latter representing the true physical conditions but not being feasible on
heavily contaminated continua.
The results indicate that modelling the pseudo continuum with a polynomial is
valid when being performed carefully with respect to the underlying absorption.
The error rises dramtically for positions near the center of the \HI absorption
which is understandable due to the low equivalent width of the \hhh feature.
However in the case of the polynomial continuum fit this effect has greater
influence. Furthermore the application of the polynomial fit to the continuum
and accordingly the rectification of the flux introduces a net shift in the
determination of the as seen in Figure
\ref{fig:compare_sim+shift}. An error that may not average out for low
statistics.
%
\subsection{Selection of lines}
The strong contamination of the observed spectrum forces us to find a balance
between a sufficiently high number of \hhh features taken into account for
better statistics and the quality of the surrounding continuum. An independent
selection via curve of growth analysis yields the same group of \hhh lines as
used in \cite{Reinhold06} and \cite{Ivanchik05}.
\section{Results}
\label{results}
\begin{figure}[h]
\begin{center}
\includegraphics[width=0.8\columnwidth]{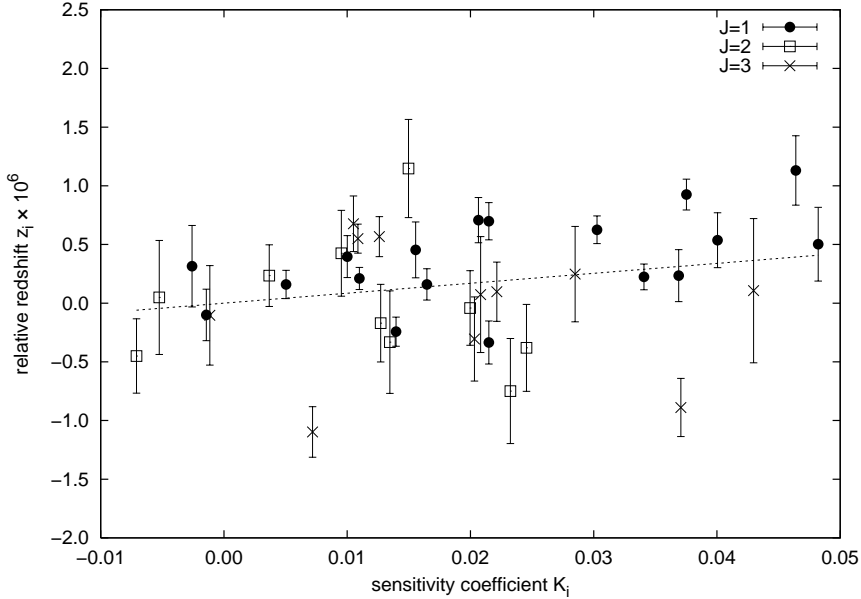}
\caption{Plot of $z_i$ -- $K_i$ relation for all observed rotational levels.
The origin of the relative redshift corresponds to $z_{\mathrm{abs}}=3.024899$.
The linear fit (\textit{dashed line}) is consistent with
$\Delta\mu/\mu=2.1\times10^{-5}$.}
\label{fig:zvsk1}
\end{center}
\end{figure}
\begin{figure}
\begin{tabular}{ccc}
\includegraphics[height=5cm,width=0.55\columnwidth]{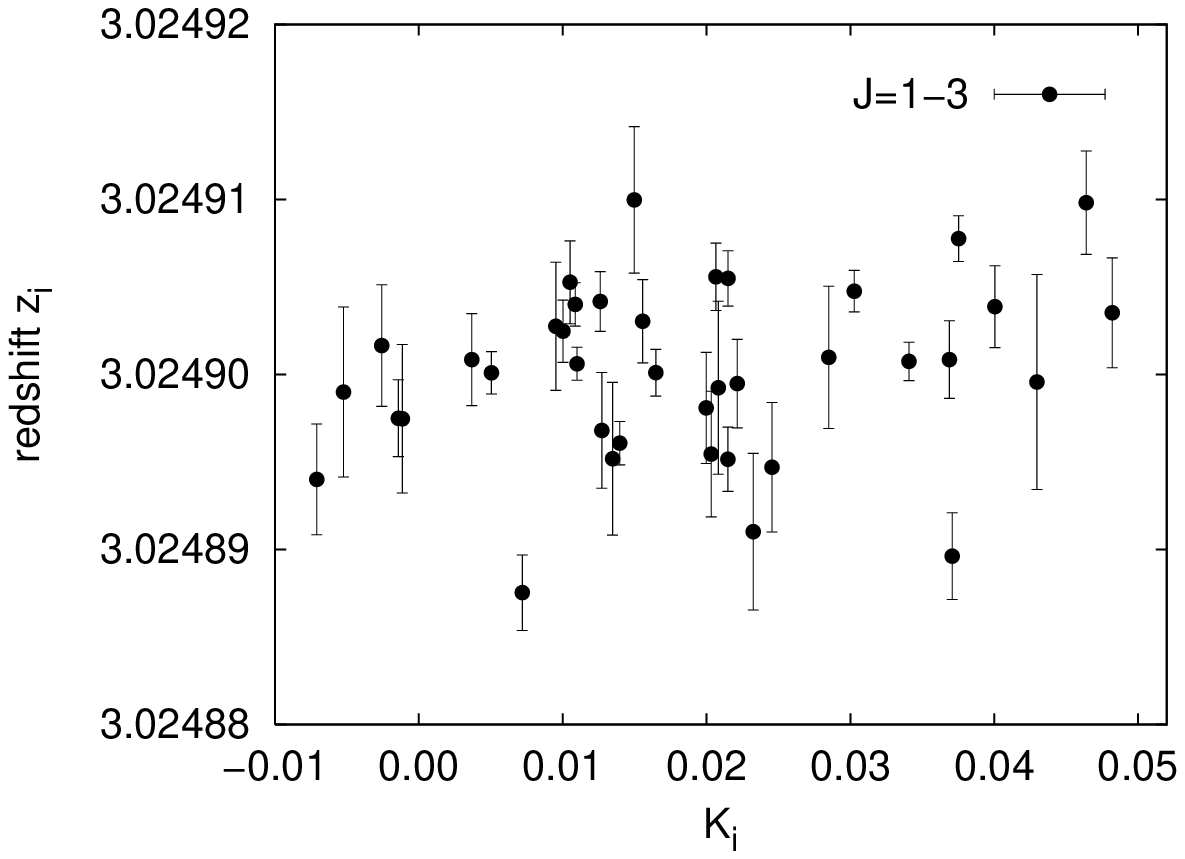}&
\hspace{-0.5cm}\includegraphics[height=5cm,width=0.45\columnwidth]
{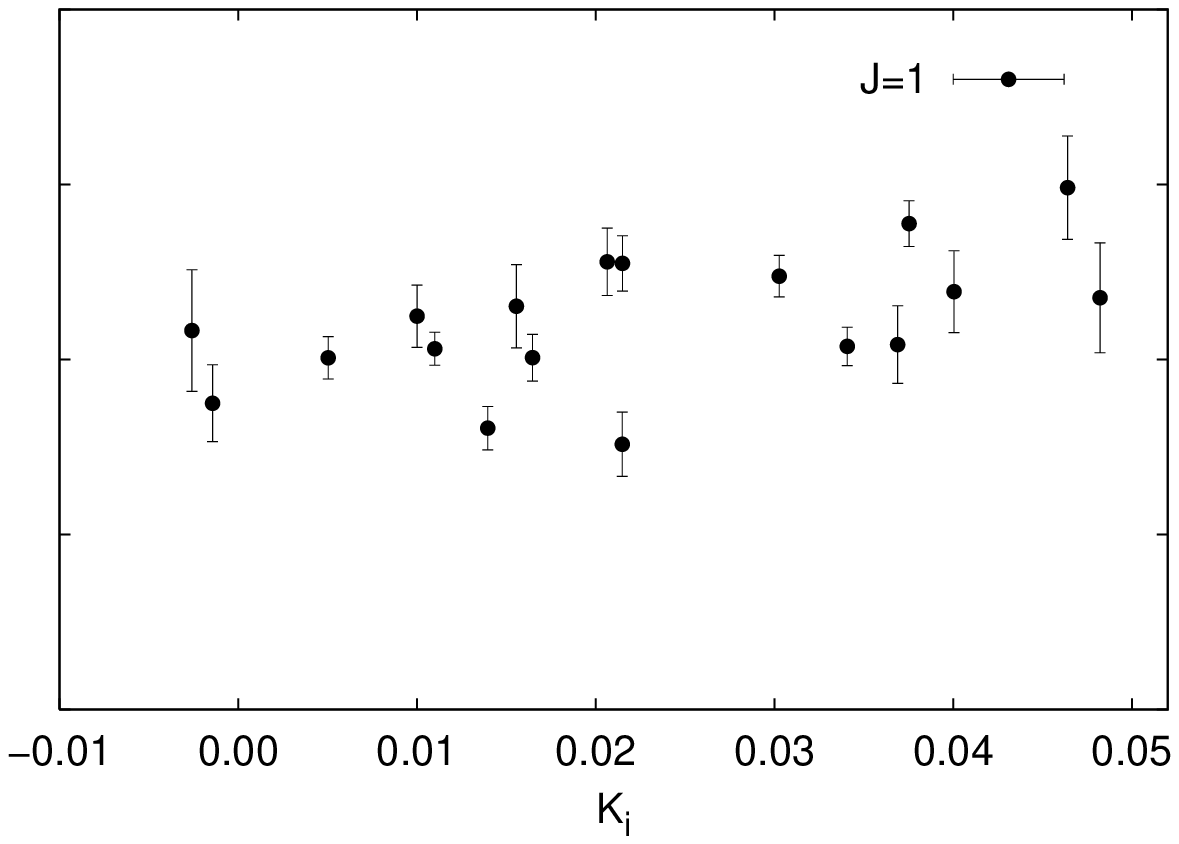}\\
\includegraphics[height=5cm,width=0.55\columnwidth]{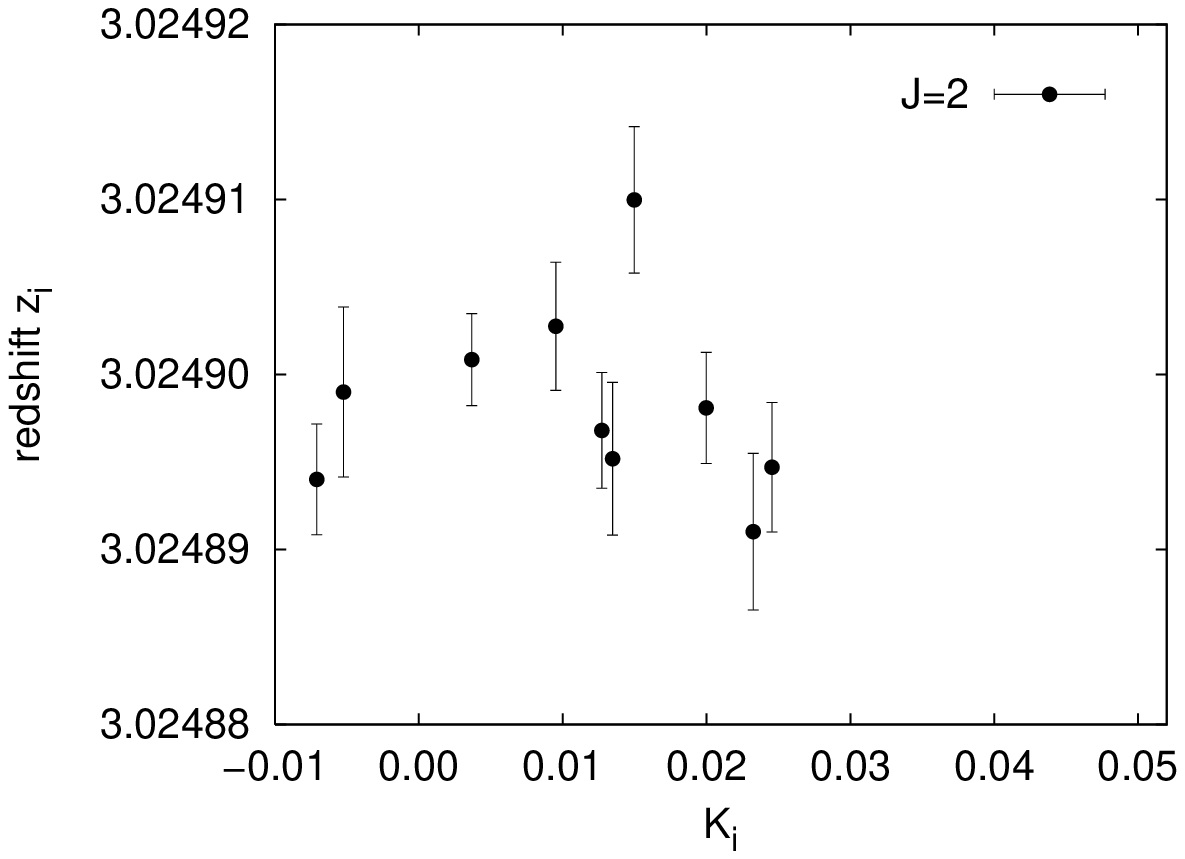}&
\hspace{-0.5cm}\includegraphics[height=5cm,width=0.45\columnwidth]
{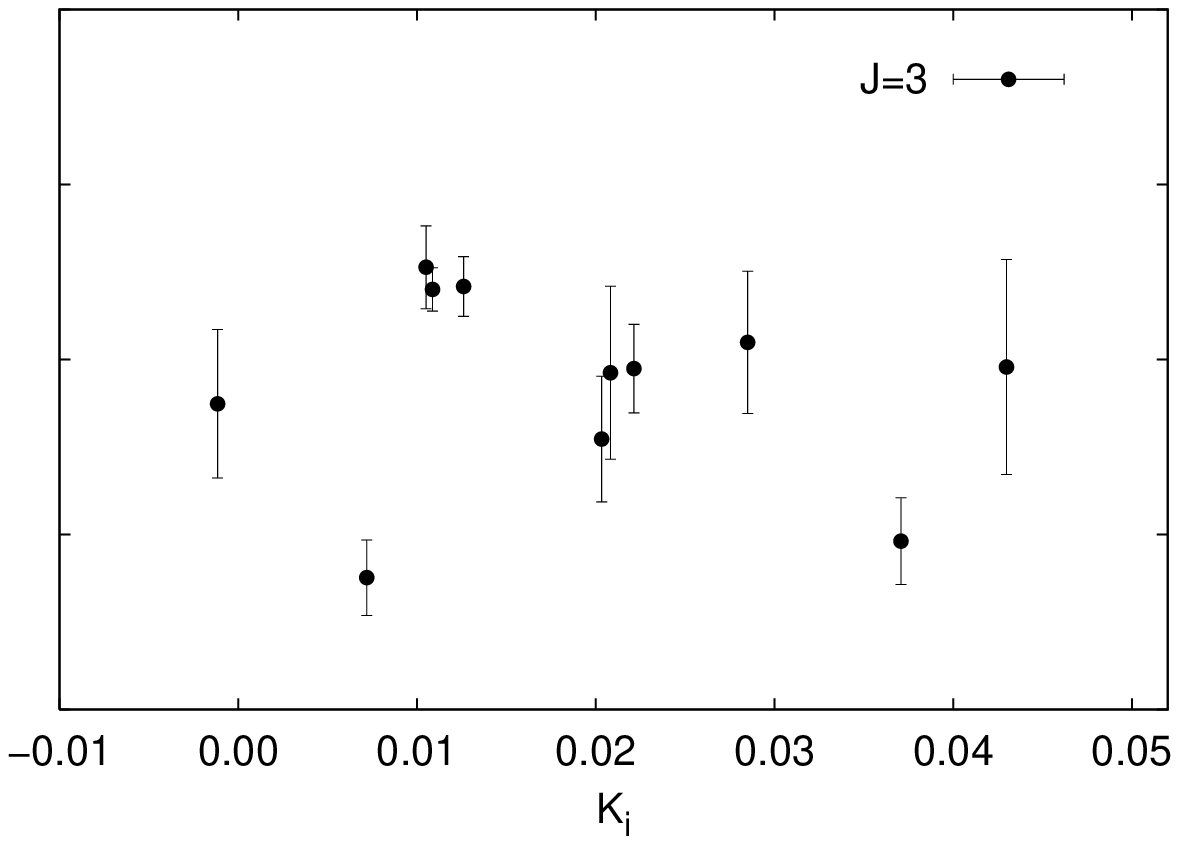}\\
\end{tabular}
\caption{$z_i$ -- $K_i$ relation with individual redshifts for the whole dataset
(\textit{top left}) and separate rotational levels J=1 to J=3 (\textit{top
right and bottom}).}
\label{fig:zvsk_jcompare}
\end{figure}

All 39 lines are fitted and a corresponding redshift is derived. The
measured redshifts with their obtained standard deviations are not consistent
with a single mean redshift at the $1\sigma$ level. 
Accordant to the method described in Section \ref{sec:1} the measured
redshifts are tested for a possible correlation with the individual sensitivity
coefficients. Figure \ref{fig:zvsk1} shows the resulting plot. The dotted line
represents a preliminary linear fit to the data.
The redshift of the absorber does not correspond to the mean
redshift, but can be ascertained by the point of zero sensitivity towards
possible $\mu$-variation, since at this point the measured redshift equals the
cosmological redshift (see Eq. \ref{eq:linear_plot}). The ordinate is given
in relative redshift with respect to the absorber
$z_{\mathrm{abs}}=3.0248990(13)$.
The linear fit corresponding to $\Delta\mu/\mu= 2.1 \pm 1.4\times10^{-5}$ is
achieved by a linear regression without taking into account the individual
estimated errors in the redshift to allow for a comparison with the results of
Reinhold \et \cite{Reinhold06} who measured $\Delta\mu/\mu= 2.06 \pm
0.79\times10^{-5}$
with the method of an unweighted fit but based on a merged dataset of two
 quasars at different redshifts.
The argumentation behind an unweighted fit is that the dispersion of the
experimental points then characterizes the true statistical errors. At a 95\%
confidence level this results gives a constraint of the variation to
$-0.7\times 10^{-5}\le\Delta\mu/\mu\le4.9\times 10^{-5}$. However, a straight
forward linear fit to the data turns out to be inappropriate.\\
Figure \ref{fig:zvsk_jcompare} illustrates the $z_i$ -- $K_i$ relation for all
three observed rotational levels combined (\textit{top left}) and each of them
separately (\textit{top right and bottom}). Evidently only the first
rotational level shows an apparent correlation between redshift $z_i$ and
sensitivity coefficients $K_i$.
Table \ref{tab:different j results} gives the mean redshifts $\bar z$ for each
corresponding rotational level. It is worth noting that the deviation
$\sigma_{\bar z}$ from
the mean value is smaller than the 2$\sigma$ level of the mean
estimated error $\bar \sigma_z$.
The number of lines is too low to make significant statistical claims.

\begin{figure}
\begin{center}
\includegraphics[width=0.75\columnwidth]{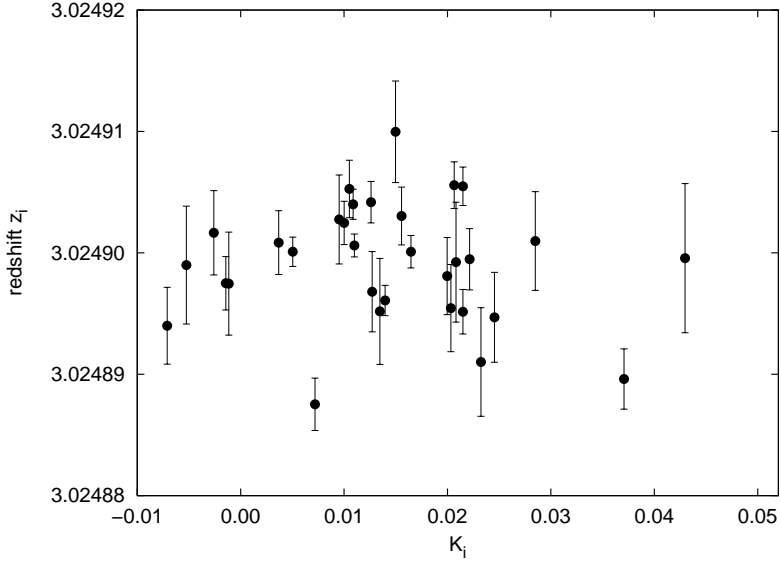}
\caption{$z_i$ -- $K_i$ relation with individual redshifts for the reduced
dataset of 32 lines in all observed rotational levels. With a Spearman
rank-order
correlation coefficient of $r_\mathrm{s}=0.08$ the data is in distinct agreement
with no correlation.}
\label{fig:subset 32 lines}
\end{center}
\end{figure}

However, the Pearson product-moment correlation coefficient (equivalent to
dividing the covariance between two variables by the product of their standard
deviations) indicates only a very weak correlation of the data. Apparently only
the first rotational level contributes to a positive linear correlation at all
(see Table \ref{tab:pearsons}). 
A closer examination of the first rotational level (see Figure
\ref{fig:zvsk_jcompare} \textit{upper right}) by eye shows that the positive
gradient of the linear
regression seems to evolve only by the rightmost seven data points.
This impression can be confirmed by a correlation test for all data points
except those seven lines with $J=1$. Table \ref{tab:pearsons} lists the
correlation coefficient of this subset as $r=-0.08$. A result quite consistent
with no correlation at all (see Fig.\ref{fig:subset 32 lines}).
However, seven data points represent a considerable subset of the total of 39
lines. The seven lines (namely
\texttt{L7R1, L8R1, L9R1, L9P1, L10P1, L13R1, L14R1}) evidently
have a sensitivity coefficient of $> 0.03$  in
common, they also emerge from high energy transitions.
As can be seen from the line identifiers, the seven lines in question all arise
from high vibrational levels in the Lyman band. This matches fully with the
range where the sensitivity coefficients deviate the most from the
Born-Oppenheimer-approximated calculations \cite{Reinhold06}.
\begin{table}[b]
\caption{Correlation coefficients for different datasets}\label{tab:pearsons}
\begin{center}
\begin{tabular}{|c|c|c|}\hline
rotational level & r & data points \\\hline\hline
1 & 0.29 & 18\\\hline
2 & -0.16 & 10\\\hline
3 & -0.62 & 11\\\hline
all & 0.24 & 39\\\hline
subset & -0.08 & 32\\\hline
\end{tabular}
\end{center}
\end{table}
%
\subsection{Goodness of fit}
To quantify the $z_i$-$K_i$ correlation further a Spearman rank-order
correlation
coefficient $r_\mathrm{s}$ is calculated. It is a non-parametric measure of
correlation, i.e., it assesses how well an arbitrary monotonic
function could describe the relationship between two variables without making
any assumptions about the frequency distribution of the variables. Unlike the
Pearson product-moment correlation coefficient, it does not require the
assumption that the relationship between the variables is linear nor does it
require the variables to be measured on interval scales.
Furthermore it allows to determine the significance of a non-zero correlation. 
Instead of measured values merely their ranks are compared.
The analysis gives coefficients of $r_\mathrm{s}=0.17$ for the full set of lines
and
$r_\mathrm{s}=0.08$ for the subset of 32 lines (see Figure \ref{fig:subset 32
lines}). The latter being not significant at the
99.9\% level (i.e., in only 0.1\% of all cases a non-zero correlation can
be deduced from a correlation coefficient of $0.08$) thus in full agreement
with
no correlation. The obtained Spearman rank-order correlation coefficients occur
in case of zero-correlation with probabilities of $\sim 1/3$ and
$2/3$,
respectively.

The goodness of fit states how well a statistical model fits a set of
observations
in contrast to a general $\sigma$ of a fit which only states the confidence of
the fit. 
The $\chi^2$ based fit allows for
a statement on the quality or
rather the likelihood of a fit in respect to the model.
Due to the simplicity of the assumed model (see Eq.\ref{eq:linear_plot}) the
relation between measured redshift and computed sensitivity coefficient is
stringently linear in case of variation. In case of no variation the gradient
would be zero but still fit the assumed linear model.
As a consequence, any result must be consistent with a linear correlation
between $z_i$ and $K_i$ independently of a possible variation.
The performed simulations of linefits indicated that the fit error on the
observed wavelength for each line and thus the error in the relative redshift
as estimated by the fitting programme are reasonable. This, of course, depends
also on the accuracy of the errors given in the observed data.
The correctness of the errors in the spectral data can be crudely tested by the 
$\chi^2$ value of a reasonably good fit. However, the $\chi^2$ is influenced by
the trueness of the error as well as the accuracy of the model. 
The mean value of all 351 fitted lines (39 lines in 9 spectra) is
$\bar{\chi^2}=1.16$ with a more descriptive median of exactly
$1.00$. The mean $\chi^2$ is expected to lie above the median due to the fact
that for some absorption features the model of a single component is evidently
wrong and thus the flux is not fitted by the model for the whole evaluated
range. 

The goodness of fit for the fitted linear relation gives a likelihood of the
fit of $\ll 0.1\%$.
This strongly indicates that the errors in general are underestimated.
A lowered likelihood can also emerge from measurement errors that are strongly
non-normally distributed. The simulations with synthesized data, however, did
show that the errors of the fit are in good agreement with a normal
distribution for the lines selected.
The validity of the assumed model is convincing and the errors of the fitted
redshifts appear to be systematically underestimated. The fitted linear relation
is not at all consistent with the observed data and the adopted error simply
reflects the statistically best solution without considering consistency with
the data.
However, the likelihood of the fit scales rather strongly with the errors of the
observed redshift. Assuming the greatest uncertainty in the measured redshifts
despite the evidently accurate output of the fitting program, the errors given
for the redshifts were scaled by a constant factor.
A scaling by a factor of two results in a likelihood of the fitted data
of about
$25\%$ which would be acceptable, though still being low.
\begin{table}[b]
\caption{The mean redshift $\bar z$ for each rotational
level, along with its standard deviation in comparison to the mean estimated
error of the redshift $\bar \sigma_z$.}\label{tab:different j results}
\begin{center}
\begin{tabular}{|c|c|c|c|}\hline
J & $\bar z$ & $\sigma_{\bar z}$ & $\bar \sigma_z$  \\\hline\hline
1 & 3.0249022 & 3.8$\times 10^{-06}$ & 1.9 $\times 10^{-06}$ \\\hline
2 & 3.0248982 & 5.3$\times 10^{-06}$ & 3.8 $\times 10^{-06}$ \\\hline
3 & 3.0248984 & 5.7$\times 10^{-06}$ & 3.2 $\times 10^{-06}$ \\\hline
\end{tabular}
\end{center}
\end{table}

Note, that for this analysis a constant shift or offset in wavelength
calibration has no impact on the deduced correlation. Only differential shifts
would influence the outcome of the analysis.
\section{Conclusions}
Asking for self consistency in the result, brings down the significance of the
measurement of a possible variation to  $\sim 1\sigma$ at once.
Rejecting the subset of seven lines of a single rotational level from the
dataset leads to no correlation at all, since the correlation is already
very weak as discussed above. A linear fit is no longer reasonable.
A detection of a positive variation of $\mu$ is caused solely from transitions
for
which $J=1$ and $v>7$ in the upper state, i.e., a range of transitions with
significant non-BOA effects as  reflected in the large changes in recently
calculated $K_i$ or $\lambda_i$ for these vibrational bands.
The discrepancy of the observed redshifts and their errors with the basic model
of $\lambda_{\mathrm{obs}}=\lambda_0\times (z_{\mathrm{abs}}+1)$ points out
yet not fully
understood systematic errors.
Possible reasons for the scatter in the redshift of several $\sigma$ can be
that a resolution of 53.000 is not sufficient for accurate linefits of
unresolved \hhh features and thus the positioning error is systematically
underestimated. 

A possible cause for differences in measured redshifts between the excitation
levels can also be a spatially separation between the points of origin. The
nature of DLAs is not
yet fully understood and subject to several inconsistencies. Recent simulations
indicate that molecular hydrogen is distributed highly inhomogeneous and clumpy
\cite{Hirashita03}.
Despite the successful refinement of transition frequencies for \hhh their
accuracy for intergalactic physical conditions requires further verification.
The yet unknown origin of the scatter of measures redshift in this work and
others demands great care in future analysis and renders the gain in
significance by combining data of several systems questionable.
It would rather indicate the need for more careful selection of analyzed
features, an in-depth study of the according transition wavelengths as well as
a high precision reduction of the observed data in all particular steps
(i.e. the data acquisition itself, wavelength calibration across several orders,
vacuum correction).
The level at which possible variations are investigated today sets high demands
on all steps involved.

A possible variation of the proton-to-electron mass ratio cannot be confirmed.
Furthermore the inconsistency of the data and the errors with respect to the
scattering of the observed redshifts give reason to be more careful in
formulating constraints as well.

The resulting constraint of this work at the 95\% level is
$|\Delta\mu/\mu|\le4.9\times
10^{-5}$ over the period of $\sim$ 11.5 Gyr, or
a maximum change of $\mu$ by $4.3 \times 10^{-15}$ yr$^{-1}$ for the
hypothetical case of
linear variation in time. The \textit{light travel time} of 11.496 Gyr for
$z=3.025$ corresponds
to the cosmological parameters $H_0 = 71\,\,\kms\mathrm{Mpc}^{-1},\,\,\Omega_M =
0.27,\,\,\Omega_{\mathrm{vac}}= 0.73$.
\section{Acknowledgements}
We are thankful for the support from the Collaborative Research Centre 676 and
for helpful comments by S.~A. Levshakov, P. Petitjean and R.~I. Thompson.

\end{document}